\documentclass[pra,amsmath,amssymb,aps,10pt,superscriptaddress,longbibliography,twocolumn]{revtex4-1}
\usepackage[utf8]{inputenc}
\usepackage{amsmath}
\usepackage{latexsym}
\usepackage{amsfonts}
\usepackage{epsfig}
\usepackage{psfrag}
\usepackage{graphicx}
\usepackage{dutchcal} 
\DeclareMathAlphabet{\mathbfcal}{U}{dutchcal}{b}{n}
\usepackage{siunitx}

\renewcommand{\vec}[1]{\boldsymbol #1}
\newcommand{\ket}[1]{\left|#1\right>}
\newcommand{\bra}[1]{\left<#1\right|}

\begin{document}
\title{Rydberg spectrum of a single trapped Ca$^+$ ion: A Floquet analysis}

\author{Mariusz Pawlak}
\email{teomar@chem.umk.pl}
\affiliation{ 
Faculty of Chemistry, Nicolaus Copernicus University in Toru\'n, Gagarina~7, 87-100~Toru\'n, Poland
}
\affiliation{
ITAMP, Harvard-Smithsonian Center for Astrophysics, Cambridge, Massachusetts 02138, USA
}
\author{H. R. Sadeghpour}
\affiliation{
ITAMP, Harvard-Smithsonian Center for Astrophysics, Cambridge, Massachusetts 02138, USA
}

\begin{abstract}
We compute the Rydberg spectrum of a single Ca$^+$ ion in a Paul trap by incorporating various internal and external coupling terms of the ion to the trap in the Hamiltonian. The coupling terms include spin--orbit coupling in Ca$^+$, charge (electron and ionic core) coupling to the radio frequency and static fields, ion--electron coupling in the Paul trap, and ion center-of-mass coupling. The electronic Rydberg states are precisely described by a one-electron model potential for e$^-$+Ca$^{2+}$, and accurate eigenenergies, quantum defect parameters, and static and tensor polarizabilities for a number of excited Rydberg states are obtained. The time-periodic rf Hamiltonian is expanded in the Floquet basis, and the trapping-field-broadened Rydberg lines are compared with recent observations of Ca$^+(23P)$ and Ca$^+(52F)$ Rydberg lines.
\end{abstract}

\maketitle

\section{Introduction}
Trapped Rydberg ions have recently come to the fore as promising candidates for fast quantum gate operations and long coherence times. The controllability and long coherence times of trapped ions, when augmented with precision and tunability of Rydberg excitations, offer tantalizing opportunities to leverage the best of the two schemes~\cite{Engel2018, Higgins_PRL2017, Higgins_PRX2017, SchmidtKaler2015}. The realization of such quantum gates with long-range Rydberg and Coulomb interactions may be used in Rydberg ion crystals for entanglement operations in quantum information processing and computing applications~\cite{Higgins_PRX2017,Zoller2008}. Additionally, trapped Rydberg ions possess enormous polarizabilities which can be manipulated with external fields~\cite{Mokhberi2019}, making them exquisite probes of their environments.

Trapped ions whose motional state fidelity is prone to decoherence due to fluctuating surface electric-field dipole noise~\cite{Hite2013,Safavi2011,Safavi2013,Lakhmanskiy2019} can be used to detect and probe residual electric fields present in Paul traps~\cite{Hite2012, Knoop2015, SchmidtKaler2015}. Rydberg atoms likewise have been shown to be sensitive probes of certain surfaces due to the presence of low electric fields~\cite{Sedlacek2016}.

The presence of static, dynamic, and stray fields in a Paul trap strongly modifies the Rydberg spectral properties of the ion. It was investigated theoretically in Refs.~\cite{Zoller2008,SchmidtKaler2011}, followed by the first realization of Rydberg $F$~\cite{SchmidtKaler2015} and $P$~\cite{Mokhberi2019} states in trapped Ca$^{+}$ ion, and the coherent control of a single trapped Rydberg Sr$^{+}$ ion in $S$ states~\cite{Higgins_PRL2017}. Accurate values for eigenenergies, transition rates, and multipole polarizabilities of low-excited states of Ca$^+$ in the absence of trapping fields were presented in Ref.~\cite{Safronova2011}.

In this work, we demonstrate details of a variational calculation of Paul-trap-induced Ca$^+(52F,23P)$ Rydberg spectra. A parametric one-electron valence potential with spin--orbit coupling is used to describe the electronic structure of bare Ca$^+$ ions. The energy levels, transition dipole moments, and quantum defects for highly excited ions are calculated for $S$, $P$, $D$, $F$, and $G$ states (up to $n=64$). The scalar and tensor polarizabilities for the Rydberg states are determined and compared with available values. A~Floquet expansion is used to calculate the matrix elements with the coupling of the ion motion to the rf and electrodes' trap potentials. The resulting Rydberg spectra with additional peaks due to the field couplings are examined and compared with recent experimental findings~\cite{SchmidtKaler2015}. Unless indicated otherwise, atomic units (a.u.) are used throughout.

\section{Theory and Computation}
\subsection{Hamiltonian terms}
The Hamiltonian of an atomic ion in a Paul trap can be written as~\cite{SchmidtKaler2011}
\begin{equation}\label{eqn_Htot}
\hat{H}=\hat{H}_{\rm e}+\hat{H}_{\rm Ie}+\hat{H}_{\rm I},
\end{equation}
where $\hat{H}_{\rm e}$ is the electronic Hamiltonian for an ion in a Paul trap, $\hat{H}_{\rm Ie}$ is the Hamiltonian describing the atomic electron coupling to the trapped ion motion, and $\hat{H}_{\rm I}$ is the Hamiltonian for the motion of the ion in the trap. Each term is described in detail below.

\subsubsection{Electron motion in the trap fields}
The coupling of the valance electron to the trapping potentials (in a.u.) is given as
\begin{equation} \label{eqn_Hp}
\hat{H}_{\rm e} = \hat H_{\rm FF}-\Phi({\vec r},t)+E_{\rm geom}z\cos(\Omega_{\rm rf}t),
\end{equation}
where the first term on the right-hand side is the field-free Hamiltonian, describing a single free Ca$^{+}$ ion. The second term is the coupling of the electron to the linear Paul trap, and the last term is the residual electric field, which the ion is exposed to due to fabrication imperfections. The magnitude of this residual field is obtained from the broadening of the $4S$--$4P$ transition in Ca$^+$~\cite{SchmidtKaler2015}. $\Omega_{\rm rf}$ is the rf frequency.

The field-free Hamiltonian is
\begin{equation}
\hat H_{\rm FF} = -\frac{1}{2}\Delta_{\vec r} + V_l(r) + V_{\rm LS}(r),
\end{equation}
where the valence electron interacts with all other electrons via an effective nonlocal parametric potential~\cite{AymarGrenne1996}:
\begin{eqnarray} \label{pot1}
V_l(r)&=&-\frac{1}{r} \left (2+(Z-2)e^{-a_1^{(l)} r}+a_2^{(l)} r e^{-a_3^{(l)} r} \right) \nonumber \\ 
&& -\frac{\alpha_c}{2r^4}\left(1-e^{-(r/r_c^{(l)} )^6} \right),
\end{eqnarray}
with $Z$ being the ion nuclear charge. The last term in Eq.~(\ref{pot1}) is the core polarization potential, wherein $\alpha_c$ is the electric dipole polarizability of the doubly charged ionic core and $r_c^{(l)}$ is a cutoff radius which ensures the proper behavior of the potential near the origin. The $l$-dependent parameters ($a_1^{(l)}$, $a_2^{(l)}$, $a_3^{(l)}$, and $r_c^{(l)}$) fitted to experimental energy levels are available for different alkaline-earth-metal ions in Ref.~\cite{AymarGrenne1996}.

The spin--orbit coupling is
\begin{equation} \label{potVLS}
V_{\rm LS}(r)=\frac{\alpha^2_{\rm LS}}{2} \frac{1}{r}\frac{{\rm d}V_l(r)}{{\rm d}r}  \left (1 -\frac{\alpha^2_{\rm LS}}{2}V_l(r)  \right )^{-2}  \hat{\textbf{L}}\cdot\hat{\textbf{S}}, 
\end{equation}
where $\alpha_{\rm LS}$ is the fine-structure constant, and $\langle \hat{\textbf{L}}\cdot\hat{\textbf{S}} \rangle =[j(j+1)-l(l+1)-3/4]/2$. The total electronic angular momentum quantum $j=l \pm \frac{1}{2}$.

The coupling of the electron to the linear Paul trap includes two terms,
\begin{equation}\label{eqn_ePhi}
\begin{aligned}
\begin{split}
-\Phi(\vec r,t)&=\hat H_{\rm trap}^{\rm rf}+\hat H_{\rm trap}^{\rm dc}\\
&=-\alpha \cos(\Omega_{\rm rf}t)(x^{2}-y^{2}) +\beta(x^{2}+y^{2}-2z^{2}),
\end{split}
\end{aligned}
\end{equation}
where $\alpha$ and $\beta$ are the rf and static electric field gradients, respectively, and $(x,y,z)$ are the electron coordinates.

Finally, the trap imperfection alternating residual electric field amplitude, see the third term in Eq.~(\ref{eqn_Hp}), is written explicitly as~\cite{SchmidtKaler2015} 
\begin{equation} \label{eqn_URF}
E_{\rm geom}=0.8(U_{\rm rf}/{\rm m})\sin(\Omega_{\rm rf}t),
\end{equation}
where $U_{\rm rf}$ is the rf voltage. The numerical coefficient in Eq.~(\ref{eqn_URF}) may be different for different ion traps.

\subsubsection{Electron motion coupled to trapped ion motion}
In a highly excited Rydberg state, the spatial extent of the electron wave function can become larger than the oscillator length, and the coupling of the electronic and external motional degrees of freedom needs to be accounted for. The Hamiltonian for the Rydberg electron coupling to the ion motion (in a.u.) is
\begin{equation} \label{eqn_HIe}
\begin{split}
\hat{H}_{\rm Ie} &=\hat H_{\rm Ie}^{\rm rf}+\hat H_{\rm Ie}^{\rm dc}\\
& = -2\alpha \cos(\Omega_{\rm rf}t)(xX-yY) +2\beta(xX+yY-2zZ),
\end{split}
\end{equation}
where $(X,Y,Z)$ are the ion coordinates in the trap.

\subsubsection{Ion motion in the trap}
The Hamiltonian for the ion center-of-mass motion in the trap is expressed as (in a.u.)
\begin{equation}\label{eqn_HI}
\hat{H}_{\rm I}=-\frac{1}{2M}\Delta_{\vec R}+\Phi(\vec R,t).
\end{equation}
The static field and the rapidly oscillating rf field form an effective time-independent harmonic potential~\cite{SchmidtKaler2011,Cook1985}, 
\begin{equation}\label{eqn_ePhi_2}
\Phi(\vec R,t)\simeq \frac{M}{2}\sum_{\rho=X,Y,Z} \omega_{\rho}^{2} \rho^{2},
\end{equation}
where $M$ is the mass of the ion and
\begin{equation}\label{eqn_omega}
\begin{aligned}
&\omega_{X}=\omega_{Y}=\sqrt{2\big[(\alpha/(M\Omega_{\rm rf}))^{2} -\beta/M\big]},\\
&\omega_{Z}=2\sqrt{\beta/M},
\end{aligned}
\end{equation}
are respectively, the transverse and axial trap frequencies.

The total Hamiltonian in Eq.~(\ref{eqn_Htot}), when grouped for computational efficiency, is 
\begin{equation}\label{eqn_Htot2}
\hat H = \hat H_{\rm FF} + [\hat H_{\rm trap}^{\rm rf} + \hat H_{\rm trap}^{\rm dc}] + \hat H_{\rm geom} + [\hat H_{\rm Ie}^{\rm rf} + \hat H_{\rm Ie}^{\rm dc} ]+ \hat H_{\rm I}.
\end{equation}

\subsection{Solutions to the field-free Hamiltonian}\label{sectionFF}
We are interested in the bound eigenstate spectrum of the stationary time-independent Schr\"{o}dinger equation describing the valence electron motion in Ca$^+$:
\begin{equation}\label{SE_FF}
\hat H_{\rm FF} \Psi = E \Psi.
\end{equation}
The bound states of this Hamiltonian are expanded in the $L^2$ basis, $\varphi_k(r)Y_{l,m}(\theta,\phi)$, where $\varphi_{k}=r^{\zeta-1} e^{-\gamma_{k}r}$ are Slater-type orbitals (STOs), with $\zeta$ and $\gamma_k$ as the optimization parameters. Because STOs are not orthogonal, we diagonalize the overlap matrix~$\mathbf{S}$: ${\vec \lambda}=\mathbf{V}^{\rm T}\mathbf{S}\mathbf{V}$, where ${\vec \lambda}$ is a diagonal matrix of positive eigenvalues and $\mathbf{V}$ is an orthogonal eigenvector matrix. Next, we transform the radial part of the basis set to an orthonormal form:
\begin{equation}
\tilde{\varphi}_p(r)=\frac{1}{\sqrt{\lambda_{p,p}}} \sum_{k} V_{k,p}\varphi_k(r).
\end{equation}
Matrix elements of the filed-free Hamiltonian in the orthonormal basis set are
\begin{equation}\label{matrixH}
\left [ \mathbf{H}_{\rm FF} \right ]_{i^\prime,i} =  \frac{\delta_{l^\prime,l} \delta_{m^\prime,m}}{\sqrt{\lambda_{p^\prime,p^\prime}\lambda_{p,p}}} \sum_{k^\prime,k} V_{k^\prime,p^\prime}^{*}  \langle \varphi_{k^\prime} |{\hat H}_{\rm FF} |\varphi_k \rangle V_{k,p}.
\end{equation}   
This variational approach allows us to calculate accurately the field-free energy spectrum for any arbitrary $(l,j)$ sets. The trial space is spanned by 660 STOs. The optimization procedure and the details for calculating the matrix elements of the Hamiltonian are provided in Ref.~\cite{Pawlak2014}. The energies resulting from the diagonalization of $\mathbf{H}_{\rm FF}$ are fully converged with respect to basis set size. The energies and wave functions also behave properly with respect to the spin--orbit splitting, which decreases with increasing orbital quantum number.

The {radial} field-free eigenfunctions are 
\begin{equation}\label{radial_eq}
\psi_n(r) = \sum_{p} \frac{C_{p,n}^{(l,j)}}{\sqrt{\lambda_{p,p}}} \sum_{k}V_{k,p}\varphi_k(r), 
\end{equation}
where the expansion coefficients $C_{p,n}^{(l,j)}$ are from the eigenvector matrix of $\mathbf{H}_{\rm FF}$.

The eigenenergies are used to determine the quantum defects $\delta_{l,j}$. Within the quantum defect theory approach, the energy levels of the system with one valence electron are given by~\cite{Drake1996,Seaton1983}
\begin{equation}\label{qdt1}
E_{n,l,j}=-\frac{Z_c^2}{2(n-\delta_{l,j}(n))^2},
\end{equation}
where $Z_c$ is the ionic core charge and $n$ is the principal quantum number. For highly excited states, it is often sufficient to take $\delta_{l,j}(n)$ as a constant. For lower excitations, the Ritz expansion is applied:
\begin{equation}\label{qdt2}
\delta_{l,j}(n)=\delta^{l,j}_0+\frac{\delta^{l,j}_2}{(n-\delta^{l,j}_0)^2}+\frac{\delta^{l,j}_4}{(n-\delta^{l,j}_0)^4} + \ldots .
\end{equation}

The scalar ($\alpha_{0}$) and tensor ($\alpha_{2}$) polarizabilities of the Ca$^{+}(52l)$ Rydberg states are calculated as~\cite{Lai2018,Khadjavi1968} 
\begin{equation}\label{polar_0}
\alpha_{0}=-\frac{2}{3}\sum_{n^\prime,l^\prime,j^\prime} (2j^\prime+1) l_{>} \begin{Bmatrix}
l & j & \frac{1}{2}\\
j^\prime & l^\prime & 1\\
\end{Bmatrix}^{2} \frac{|\langle nl|r|n^\prime l^\prime \rangle|^2}{E_{n,l,j}-E_{n^\prime,l^\prime,j^\prime}},
\end{equation}

\begin{equation}\label{polar_2}
\begin{split}
\alpha_{2}=&-2 \sqrt{\frac{10j(2j-1)(2j+1)}{3(j+1)(2j+3)}} \sum_{n^\prime,l^\prime,j^\prime}(-1)^{j+j^\prime}(2j^\prime+1)l_{>} \\
& \times \begin{Bmatrix}
l & j & \frac{1}{2}\\
j^\prime & l^\prime & 1\\
\end{Bmatrix}^{2} \begin{Bmatrix}
j & j^\prime & 1\\
1 & 2 & j\\
\end{Bmatrix}\frac{|\langle n l|r|n^\prime l^\prime \rangle|^2}{E_{n,l,j}-E_{n^\prime,l^\prime,j^\prime}},
\end{split}
\end{equation}
where $l_{>}$ is the greater of $l$ and $l^\prime$. The total polarizability of a state with non-zero total angular momentum is~\cite{Kamenski2014}
\begin{equation}\label{polar_tot}
\alpha_{\rm tot}=\alpha_{0}+\alpha_{2} \frac{3m_{j}^{2}-j(j+1)}{j(2j-1)},
\end{equation}
with $-j\leq m_j\leq j$.

\subsection{Floquet solutions to the time-periodic Hamiltonian}
Since the Hamiltonian is time periodic, we expand the solutions in a Floquet basis~\cite{Shirley1965,Moiseyev1990,Moiseyev1991,Moiseyev1994,MoiseyevPRA1994}, leading to the eigenvalue equation (in a.u.)
\begin{equation}
\hat {\cal H}_{\cal F} {{\cal Y}_\varepsilon (\vec R,\vec r,t)} = \left (\hat H- i\frac{\partial}{\partial t} \right ) {{\cal Y}_\varepsilon (\vec R,\vec r,t)} =\varepsilon {\cal Y}_\varepsilon (\vec R,\vec r,t),
\end{equation}
where $\hat {\cal H}_{\cal F}$ is the Floquet Hamiltonian and ${\cal Y}_\varepsilon (\vec r,\vec R,t)$ are time-periodic wave functions with period $2\pi/\Omega_{\rm rf}$,
\begin{equation}
{\cal Y}_\varepsilon (\vec r,\vec R,t) = {\cal Y}_\varepsilon \left (\vec r,\vec R,t+\frac{2\pi}{\Omega_{\rm rf}} \right ) = \sum_{q=-\infty}^{\infty} e^{iq\Omega_{\rm rf}t} \Xi_\varepsilon^q(\vec r,\vec R).
\end{equation} 
The time-independent components $\Xi_\varepsilon^q(\vec r,\vec R)$ are usually called the Floquet channel functions. They fulfill the relationship
\begin{equation}
\Xi_\varepsilon^q(\vec r,\vec R)=\Xi_{\varepsilon+g\Omega_{\rm rf}}^{q+g}(\vec r,\vec R)
\end{equation}
for any integer $g$. We represent each component in the basis set 
\begin{equation} \label{basis_set_xi}
\xi_\eta(\vec r,\vec R)= \psi_n(r) Y_{l,m}({\theta,\phi}) \prod_{\rho=X,Y,Z} \psi_{k_\rho}(\rho),
\end{equation} 
{where $\eta$ is a superindex containing the quantum numbers $\{n,l,m,k_X,k_Y,k_Z\}$.}
Solutions of a three-dimensional quantum harmonic oscillator, $\Pi_{\rho=X,Y,Z}\psi_{k_\rho}(\rho)$, are used here as a part of the basis set, since by Eq.~(\ref{eqn_ePhi_2}) 
\begin{eqnarray}\label{eq:motional}
 \left \langle \prod_{\rho=X,Y,Z} \psi_{k_{\rho}^{\prime}} \left |\hat H_{\rm I} \right | \prod_{\rho=X,Y,Z} \psi_{k_\rho} \right \rangle  \nonumber \\
 =\sum_{\rho=X,Y,Z} \omega_{\rho} \left(k_\rho+\frac{1}{2} \right) \delta_{k_{\rho}^{\prime}k_\rho},
\end{eqnarray}
where $k_{\rho}$ are the harmonic oscillator quantum numbers.

Then, the Floquet--Hamiltonian matrix is expressed as
\begin{eqnarray}
\left [ \mathbfcal{H}_{\cal{F}}\right ]_{q',\eta',q,\eta} &=& \frac{\Omega_{\rm rf}}{2\pi} \int\limits_0^{2\pi/\Omega_{\rm rf}}    \langle  \xi_{\eta^\prime}(\vec r,\vec R) |\hat H(\vec r,\vec R, t)| \xi_{\eta}(\vec r,\vec R) \rangle \nonumber \\ 
&& \times \, e^{i(q-q^\prime)\Omega_{\rm rf} t}  dt +  q\Omega_{\rm rf} \delta_{q',q} \delta_{\eta',\eta}. 
\end{eqnarray}
Since the rf field in the Paul trap is sinusoidal, the Floquet--Hamiltonian matrix is reduced to the following form:
\begin{eqnarray}
\left [ \mathbfcal{H}_{\cal{F}}\right ]_{q',q}&=&\left [  \mathbf{E} + \mathbf{H}_{\rm trap}^{\rm dc}  + \mathbf{H}_{\rm Ie}^{\rm dc} + \mathbf{H}_{\rm I} +   q\Omega_{\rm rf}  \mathbf{I} \right ] \delta_{q',q} \nonumber \\
&&+\frac{1}{2}\left [ \mathbf{H}_{\rm trap}^{\rm rf} + \mathbf{H}_{\rm Ie}^{\rm rf} +\mathbf{H}_{\rm geom} \right ] \delta_{q',q\pm 1}. \label{Floquet-Matrix}
\end{eqnarray}

The explicit expressions for the matrix elements are given in the Appendix. Analytical solutions for all the angular matrix elements are reported in the Supporting Information of Ref.~\cite{Pawlak2017}. To construct the supermatrix in Eq.~(\ref{Floquet-Matrix}), the Ca$^+(n\leq 64, l\leq 4, |m|\leq l, {j=l-1/2})$ 12 photon absorption and emission transitions are considered ($q=-12,-11,...,11,12$). Since the calculations are time-consuming, the ion is assumed to be in the ground motional state ($k_X=k_Y=k_Z=0$). This approximation is physically motivated, since the mass of the ionic core is much larger than the mass of the valence electron.

Equation~(\ref{Floquet-Matrix}) can be, in general, presented in matrix form as
\begin{widetext}
\begin{equation}\label{mega_matrix}
\mathbfcal{H}_{\cal{F}} =
\left [
\begin{tabular}{@{\quad}c@{\quad}c@{\quad}c@{\quad}c@{\quad}c@{\quad}c@{\quad}c@{\quad}}
$\ddots$ & $\ddots$    &   &   &   &   &  \\
$\ddots$ & $\mathbf{A} -2 \Omega_{\rm rf} \mathbf{I}$ & $\mathbfcal{V}$ &   &   &   &  \\
         & $\mathbfcal{V}^\dag$ & $\mathbf{A} -  \Omega_{\rm rf} \mathbf{I}$ & $\mathbfcal{V}$ &   &   &  \\
         &   & $\mathbfcal{V}^\dag$ & $\mathbf{A}$  & $\mathbfcal{V}$ &   &  \\
         &   &   & $\mathbfcal{V}^\dag$ & $\mathbf{A}  + \Omega_{\rm rf} \mathbf{I}$ & $\mathbfcal{V}$ & \\
         &   &   &   & $\mathbfcal{V}^\dag$ & $\mathbf{A} +2  \Omega_{\rm rf} \mathbf{I}$ & $\ddots$  \\
         &   &   &   &   &  $\ddots$  &  $\ddots$  
\end{tabular}  
\right ]
\end{equation}
\end{widetext}
with
\begin{eqnarray}
&& \mathbf{A} = \mathbf{E} + \mathbf{H}_{\rm trap}^{\rm dc} + \mathbf{H}_{\rm Ie}^{\rm dc} + \mathbf{H}_{\rm I},   \\
&& \mathbfcal{V} =  \frac{1}{2}\left( \mathbf{H}_{\rm trap}^{\rm rf} + \mathbf{H}_{\rm Ie}^{\rm rf} +\mathbf{H}_{\rm geom} \right).
\end{eqnarray}
The convergence of the Floquet approach is examined by including progressively more Floquet basis sets (5, 9, 15, 21, and 25 channels). {Our final matrix is prepared for the following quantum numbers: $n\le 64$, $l\le 4$, $|m|\le l$, $j=l-\frac{1}{2}$, $k_X=0$, $k_Y=0$, $k_Z=0$, and $|q|\le 12$. This yields the matrix with the size of 38025 $\times$ 38025 to be diagonalized.}

\subsection{Stark effect} 
To calculate energy spectrum of the ion interacting with an external homogeneous static electric field and prepare a Stark map, we consider the Hamiltonian in the form (in a.u.)
\begin{equation}\label{Stark}
\hat H^{\rm dc}_{\rm e} = {\hat H}_{\rm FF}+E_{\rm dc}z, 
\end{equation}
where $E_{\rm dc}$ is the dc field strength. The field is chosen along the $z$ direction. We variationally solved the Schr\"odinger equation using STOs of different spherical symmetry ($l=0,1,2,3,4$). The matrix elements of the perturbed term of Eq.~(\ref{Stark}) do not vanish for $l^\prime=l\pm 1$. Since $m$ is a good quantum number here, we took $m=0$ for which the Stark effect is the largest. A~similar computational approach was applied in Ref.~\cite{Pawlak2011}.

\subsection{Oscillator strength} 
The intensity of transitions between states
${\cal Y}_{\nu^{\prime}}$ and ${\cal Y}_{\nu}$ is given by the oscillator strength (in a.u.):
\begin{equation}
f_{\nu',\nu}=\frac{2}{3}\omega_{\nu',\nu} \left | \mu_{\nu^\prime,\nu} \right |^2, 
\end{equation}
where $\omega_{\nu',\nu}=\varepsilon_{\nu}-\varepsilon_{\nu^\prime}$ is the energy difference between two states.
The transition dipole moment for the singly charged ion in a Paul trap, where all effects described above are included, reads  
\begin{eqnarray}
\mu_{\nu^\prime,\nu} &=& \langle {\cal Y}_{\nu^\prime} |r \cos\theta| {\cal Y}_{\nu} \rangle \nonumber \\
& = & \sum_{\eta^\prime,\eta}  D^{*}_{q^\prime=0,n^\prime,l^\prime,m^\prime, k_X^\prime, k_Y^\prime,k_Z^\prime;\nu^\prime} D_{q=0,n,l,m, k_X, k_Y,k_Z;\nu} \nonumber \\
&& \times \sum_{p^\prime,p} \left ( C_{p^\prime,n^\prime}^{(l^\prime,j^\prime)} \right )^{*} C_{p,n}^{(l,j)} \langle  {\tilde{\varphi}}_{p^\prime} |r| {\tilde{\varphi}}_{p}  \rangle \langle Y_{l^\prime,m^\prime} |\cos\theta| Y_{l,m}\rangle \nonumber \\ && \times \delta_{k_X^\prime,k_X} \delta_{k_Y^\prime,k_Y} \delta_{k_Z^\prime,k_Z}, \label{osc_str_eq}
\end{eqnarray}
where $D_{q,n,l,m, k_X, k_Y,k_Z;\nu}$ are the expansion coefficients obtained by diagonalizing the Floquet--Hamiltonian matrix [Eq.~(\ref{mega_matrix})]. We chose the middle Floquet channel that provides the most reliable spectrum, hence $q=0$. In the computations of the oscillator  strengths for $3D_{3/2} \rightarrow 52F_{5/2}$ and $3D_{3/2} \rightarrow 23P_{1/2}$ transitions, we put in Eq.~(\ref{osc_str_eq}) $n^\prime=3$, $l^\prime=2$, $n=52$, and $l=3$ and 
$n^\prime=3$, $l^\prime=2$, $n=23$, and $l=1$, respectively.

\section{Results and Discussion}
\subsection{Rydberg energies, quantum defects, and scalar and tensor polarizabilities}
The low-lying $nD_{3/2}$ and high-lying $nF_{5/2}$ and $nP_{1/2}$ energy levels as well as the corresponding transition dipole moments are presented in Table~\ref{tab_eng}; we calculate all Ca$^+(n\leq 64,l\leq 4)$ eigen-energies and eigen-functions. We find good agreement with the results reported by Djerad in Ref.~\cite{Djerad1991} based on the quantum defect theory and experimental data taken from Ref.~\cite{Moore1971}. Furthermore, our results exhibit even better agreement with recent experimental findings of Mokhberi and coworkers~\cite{Mokhberi2019}, especially for the highly excited states. The energy levels of Refs.~\cite{Djerad1991,Mokhberi2019}, listed in Table~\ref{tab_eng}, are determined using quantum defect parameters and expressions provided by the authors of these two references. One prominent feature with our results is that, although the basis set is generally optimized for highly excited Rydberg states, the accuracy in the low-energy states is still maintained.

\begin{table}
\caption {\label{tab_eng}
The low-lying $nD_{3/2}$ and high-lying $nF_{5/2}$ and $nP_{1/2}$ energy levels of Ca$^{+}$. The presented results are compared with Ref.~\cite{Djerad1991}, where the spin--orbit splitting is neglected, and with Ref.~\cite{Mokhberi2019}. The calculated transition dipole moments, $d_{DF}=\bra{3D_{3/2}}r\ket{nF_{5/2}}$ and $d_{DP}=\bra{3D_{3/2}}r\ket{nP_{1/2}}$, are given in the last column. The values are in atomic units.}
\begin{center}
\begin{tabular}{@{}rcccc@{}}
\hline \hline
& \multicolumn{1}{c}{$E_{nD}(\times 10^{-1})$}  &  & \multicolumn{1}{c}{$E_{nD_{3/2}}(\times 10^{-1})$} &  \multicolumn{1}{c}{} \\
$n$\, & \multicolumn{1}{c}{Ref.~\cite{Djerad1991}}  &  & \multicolumn{1}{c}{This work} &  \\
\cline{1-5}
 3 & $-3.73917$ \:   & & $-$3.74136 \:    \\
 4 & $-1.77235$ \:   & & $-$1.77338 \:    \\
 5 & $-1.04894$ \:   & & $-$1.04878 \:    \\
 6 & $-0.693570$     & & $-$0.693448    \\
 7 & $-0.492592$     & & $-$0.492543    \\
 8 & $-0.367889$     & & $-$0.367876    \\
 9 & $-0.285204$     & & $-$0.285207    \\
10 & $-0.227570$     & & $-$0.227579    \\
\\
 & \multicolumn{1}{c}{$E_{nF}(\times 10^{-3})$}  &  \multicolumn{1}{c}{$E_{nF_{5/2}}(\times 10^{-3})$}  & \multicolumn{1}{c}{$E_{nF_{5/2}}(\times 10^{-3})$} &   \multicolumn{1}{c}{$d_{DF}(\times 10^{-2})$} \\
$n$\,  & \multicolumn{1}{c}{Ref.~\cite{Djerad1991}}  &  \multicolumn{1}{c}{Ref.~\cite{Mokhberi2019}}  &  \multicolumn{1}{c}{This work} &  \multicolumn{1}{c}{This work} \\
\hline
45 & $-0.988791$ & $-0.988929$ & $-0.988967$ &  2.38399 \\
46 & $-0.946244$ & $-0.946373$ & $-0.946408$ &  2.30637 \\
47 & $-0.906385$ & $-0.906505$ & $-0.906539$ &  2.23288 \\
48 & $-0.868992$ & $-0.869106$ & $-0.869137$ &  2.16321 \\
49 & $-0.833867$ & $-0.833973$ & $-0.834003$ &  2.09709 \\
50 & $-0.800829$ & $-0.800929$ & $-0.800957$ &  2.03428 \\
51 & $-0.769716$ & $-0.769810$ & $-0.769837$ &  1.97457 \\
52 & $-0.740382$ & $-0.740471$ & $-0.740496$ &  1.91769 \\
53 & $-0.712693$ & $-0.712777$ & $-0.712801$ &  1.86350 \\
54 & $-0.686529$ & $-0.686609$ & $-0.686631$ &  1.81182 \\
55 & $-0.661780$ & $-0.661855$ & $-0.661876$ &  1.76249 \\
\\
 & \multicolumn{1}{c}{$E_{nP}(\times 10^{-2})$}  &  \multicolumn{1}{c}{$E_{nP_{1/2}}(\times 10^{-2})$} & \multicolumn{1}{c}{$E_{nP_{1/2}}(\times 10^{-2})$} &    \multicolumn{1}{c}{$d_{DP}(\times 10^{-3})$} \\
$n$\,  & \multicolumn{1}{c}{Ref.~\cite{Djerad1991}}  &  \multicolumn{1}{c}{Ref.~\cite{Mokhberi2019}} &  \multicolumn{1}{c}{This work} &  \multicolumn{1}{c}{This work} \\
\hline
20 & $-0.580433$ & $-0.580421$ & $-0.580590$  &   8.07380 \\
21 & $-0.522603$ & $-0.522596$ & $-0.522737$  &   7.45575 \\ 
22 & $-0.473006$ & $-0.473002$ & $-0.473120$  &   6.91301 \\
23 & $-0.430148$ & $-0.430147$ & $-0.430247$  &   6.43329 \\ 
24 & $-0.392862$ & $-0.392862$ & $-0.392948$  &   6.00677 \\ 
25 & $-0.360222$ & $-0.360223$ & $-0.360298$  &   5.62551 \\ 
26 & $-0.331487$ & $-0.331489$ & $-0.331554$  &   5.28306 \\ 
27 & $-0.306059$ & $-0.306061$ & $-0.306118$  &   4.97408 \\ 
28 & $-0.283448$ & $-0.283450$ & $-0.283500$  &   4.69415 \\ 
29 & $-0.263253$ & $-0.263256$ & $-0.263300$  &   4.43958 \\ 
30 & $-0.245142$ & $-0.245145$ & $-0.245184$  &   4.20727 \\
\hline \hline
\end{tabular}
\end{center}
\end{table}

The quantum defect parameters in Eq.~(\ref{qdt2}), are presented in Table~\ref{tab_qdt}. All calculated energy levels are used to fit to parameters in Eq.~(\ref{qdt1}). The recently experimentally extracted quantum defect values $\delta_0^{l,j}$ for Ca$^+(nP_{1/2})$ and Ca$^+(nF_{5/2})$ states are, respectively, $1.43690(3)$ and $0.02902(2)$~\cite{Mokhberi2019}. The goodness of our fit, within a nonlinear least-squares procedure, is as follows: the sum of squares due to error, also known as the sum of squares of residuals, is less than $2.2\times 10^{-9}$, the root-mean-squared error is less than $6.1\times 10^{-6}$, whereas the coefficient of determination ($R^2$) for the worst case is equal to one with an accuracy to seven decimal places.

\begin{table}[h]
\caption {\label{tab_qdt} Calculated quantum defect parameters in Eq.~(\ref{qdt2}) for Ca$^{+}$ in different $S$, $P$, $D$, $F$, and $G$ states.}
\begin{tabular}{l@{\quad}l@{\quad}l@{\quad}l}
\hline \hline
Level & \multicolumn{1}{c}{$\delta^{l,j}_0$} &  \multicolumn{1}{c}{$\delta^{l,j}_2$} &  \multicolumn{1}{c}{$\delta^{l,j}_4$} \\
\hline
$nS_{1/2}$ & 1.80149622    & \,\: 0.201535974  & 0.312279201  \\
$nP_{1/2}$ & 1.43927290    & \,\: 0.331987211  & 0.687628538  \\
$nP_{3/2}$ & 1.43532329    & \,\: 0.332803651  & 0.690740476  \\
$nD_{3/2}$ & 0.627759022   &  $-0.0148289072$  & 1.98904443   \\
$nD_{5/2}$ & 0.627066817   &  $-0.0128801411$  & 1.96257423   \\
$nF_{5/2}$ & 0.0298974503  &  $-0.202650265$   & 0.497416258  \\
$nF_{7/2}$ & 0.0296853803  &  $-0.198110426$   & 0.457337917  \\
$nG_{7/2}$ & 0.00614904531 &  $-0.0419102731$  & 0.0164937590 \\
$nG_{9/2}$ & 0.00614352002 &  $-0.0418535034$  & 0.0717188390 \\
\hline \hline
\end{tabular}
\end{table}

\begin{table}
\caption {\label{tab_polar} The scalar, $\alpha_{0}$, and tensor, $\alpha_{2}$, polarizabilities of the Rydberg states of Ca$^{+}$. All the values are in MHz/(V/cm)$^2$.}
\begin{center}
\begin{tabular}{l@{\quad}rrc@{\quad}rr}
\hline \hline
 &  \multicolumn{2}{c}{$\alpha_{0}$}   &  &  \multicolumn{2}{c}{$\alpha_{2}$} \\
\cline{2-3}
\cline{5-6}
Level & \multicolumn{1}{c}{This work} &  \multicolumn{1}{c}{Ref.~\cite{Kamenski2014}}& & \multicolumn{1}{c}{This work} &  \multicolumn{1}{c}{Ref.~\cite{Kamenski2014}} \\
\hline
52$P$       &           & $-69.774$  & &          &  15.137   \\
52$P_{1/2}$ & $-69.836$ &            & &   0.000  &           \\
52$P_{3/2}$ & $-68.519$ &            & &  14.980  &           \\
\\
52$D$       &          &  32.149  & &            & $-39.221$ \\
52$D_{3/2}$ &  31.857  &          & & $-27.559$  &           \\
52$D_{5/2}$ &  31.260  &          & & $-38.421$  &           \\
\\
52$F$       &          &    \multicolumn{1}{c}{---}   & &    &   \multicolumn{1}{c}{---}    \\
52$F_{5/2}$ & 582.449  &          & & $-212.689$ &          \\  
52$F_{7/2}$ & 587.395  &          & & $-250.175$ &          \\
\hline \hline
\end{tabular}
\end{center}
\end{table}

\begin{figure}[h!]
\includegraphics[width=8.4cm]{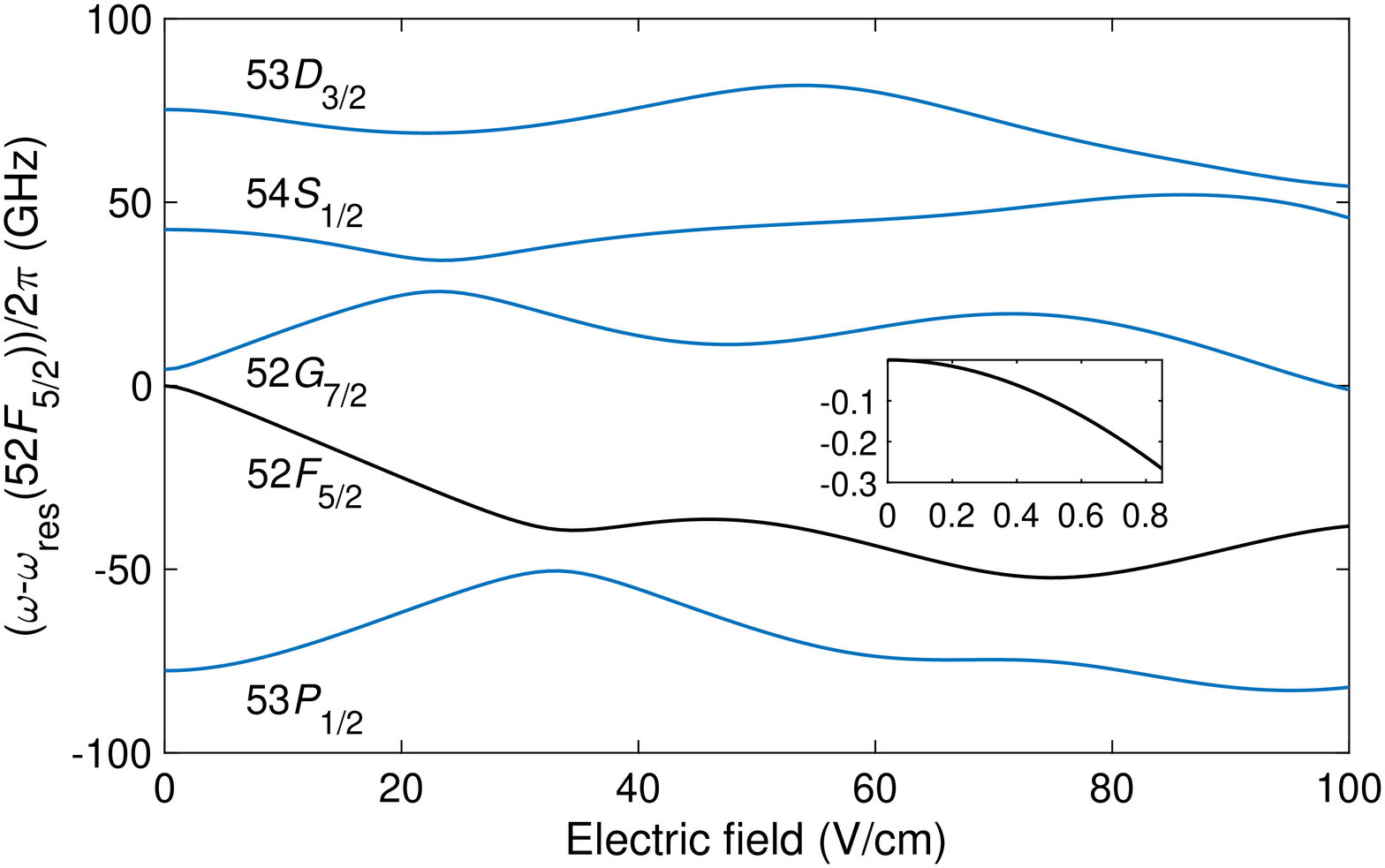}
\caption{\label{fig:Stark} 
Stark map of Ca$^{+}$ near the 52$F$ level showing the mixing of Rydberg states with different angular momenta in an electric field. As the inset indicates, at small fields, relevant to experimental values, the 52$F$-state energy shift is quadratic in the field and there is no field mixing.}
\end{figure}

A good measure of the accuracy of our wave functions and energies are the static and tensor polarizabilities for Rydberg states, as the polarizability is an acutely sensitive parameter of the linear response theory to perturbations by external fields. The static scalar and tensor polarizabilities are defined in Eqs.~(\ref{polar_0}) and (\ref{polar_2}). We calculate the Ca$^+(n=52)$ Rydberg polarizabilities, presented in Table~\ref{tab_polar}, and compared with available values from literature~\cite{Kamenski2014}. The summation in Eqs.~(\ref{polar_0}) and (\ref{polar_2}) is performed over bound states up to $n^\prime=64$. The experimentally determined $\alpha_{\rm tot}({52F})=10^{+7}_{-3}\times 10^{2}$~MHz/(V/cm)$^{2}$~\cite{SchmidtKaler2015,footnote_about_51F} is in agreement with the theoretical results in Table~\ref{tab_polar}, i.e., $\alpha_{\rm tot}({52F_{5/2}})=752.600$ and $\alpha_{\rm tot}({52F_{7/2}})=766.091$~MHz/(V/cm)$^{2}$.

\begin{figure}
\includegraphics[width=8.4cm]{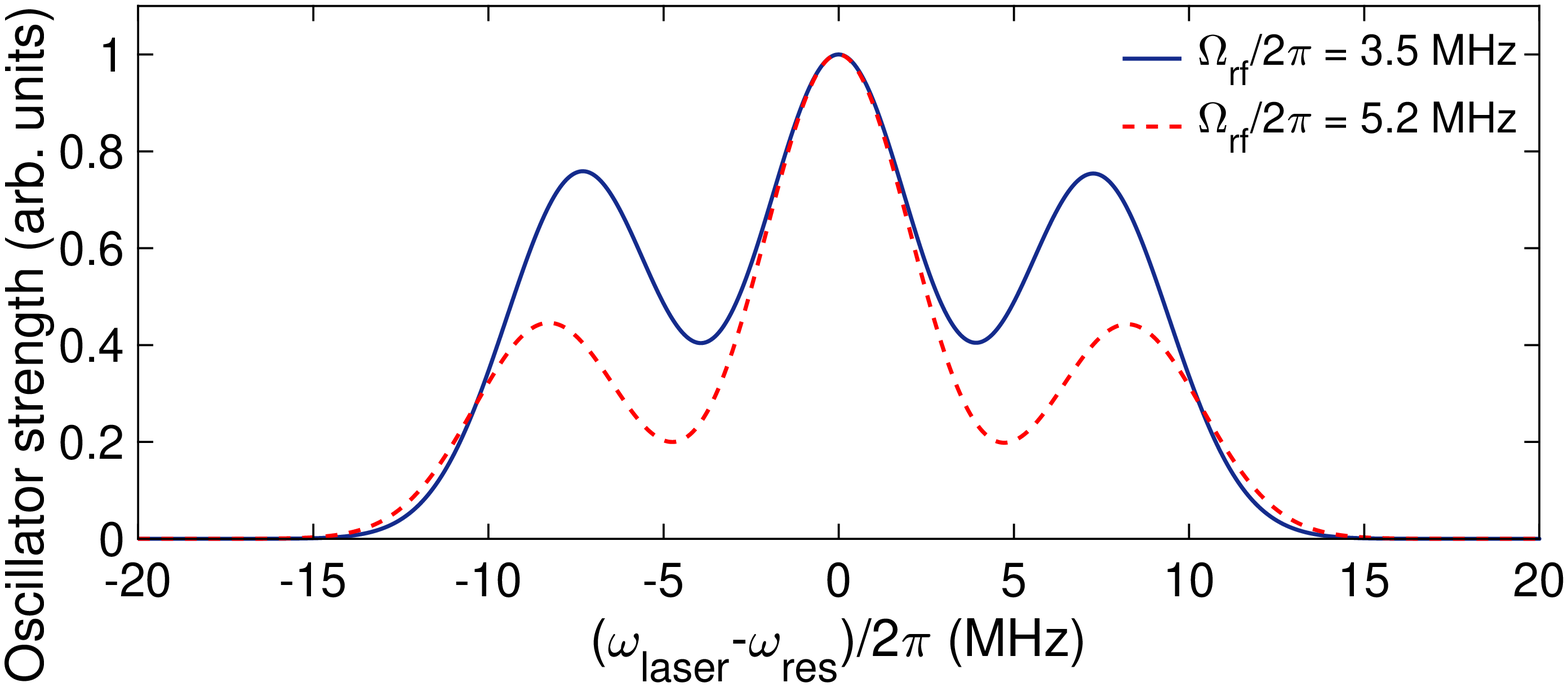}\\
\vspace{0.2cm}
\includegraphics[width=8.4cm]{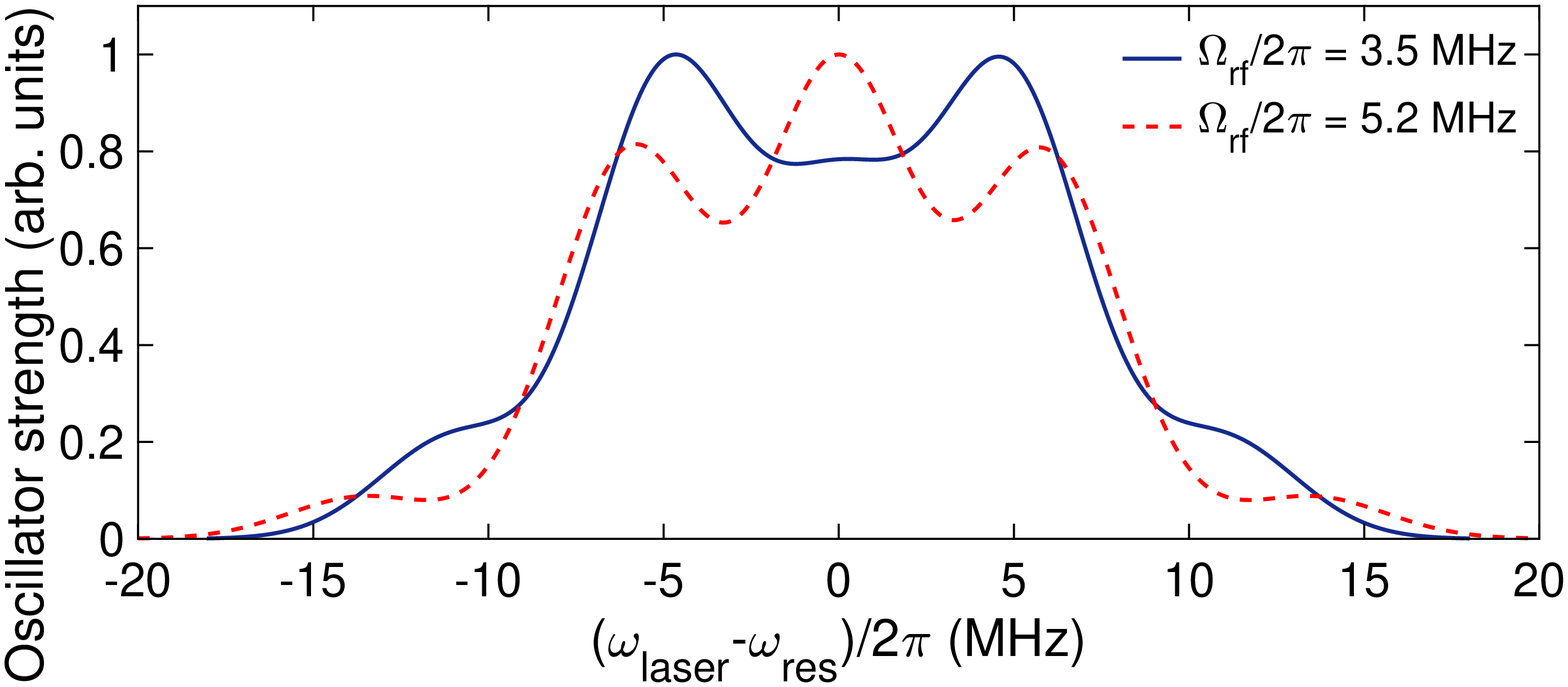}\\
\vspace{0.2cm}
\includegraphics[width=8.4cm]{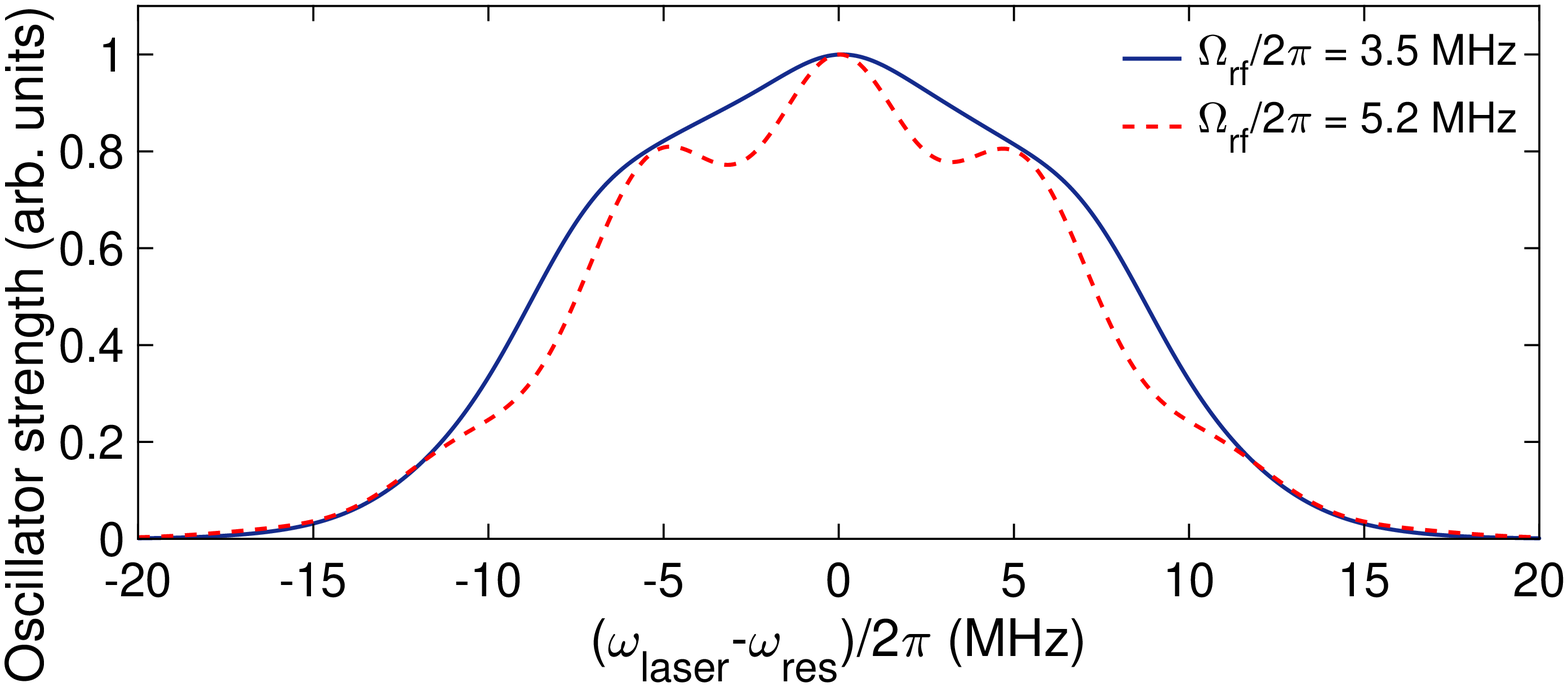}
\caption{ \label{fig:Mollow_Triplet} 
The Mollow triplet effect around the zero detuning of $3D_{3/2}$ to $52F_{5/2}$ transition line, when $\Omega_{\rm rf}/2\pi=3.5$ (solid curve) and 5.2~(dashed curve)~MHz. The rf and static field gradients are $\alpha=8.52\times 10^{6}$~V/m$^{2}$ and $\beta=3.32\times 10^{4}$~V/m$^{2}$, respectively. The residual electric field is not considered in these calculations, i.e., $E_{\rm geom}=0$. Upper panel: The calculations limited to 1-photon processes, by including 3 Floquet channels, i.e., $(-1,0,+1)$. Middle panel: The calculations limited to 2-photon processes (5 Floquet channels). Lower panel: Up to 12-photon processes allowed (25 Floquet channels). The Gaussian convolution is performed on the calculated results by considering a~5-MHz laser linewidth.} 
\end{figure}

Polarizability is proportional to the squares of transition dipole moments and inversely proportional to the energy differences [see Eqs.~(\ref{polar_0}) and (\ref{polar_2})], and thus the polarizability of the \textit{nF} states is significantly larger in comparison with the polarizability for the separated states with not-negligible quantum defects ($l<3$ states in Table~\ref{tab_qdt})~\cite{Smirnov_book}. The main contribution to the polarizability of the 52$F$ state comes from the coupling to the nearby 52$G$ state. Figure~\ref{fig:Stark} shows the Stark map of Ca$^{+}$ eigenstates in the vicinity of the 52$F$ state up to 100~V/cm. Electric fields in the ion traps are usually less than 1~V/cm~\cite{Feldker_Thesis}; at such low electric fields, the $F$ state is well isolated from the $G$ state. As expected, the inset of Fig.~\ref{fig:Stark} shows that for small fields the energy shift remains quadratic and there is no field mixing.

\begin{figure}
\includegraphics[width=\columnwidth]{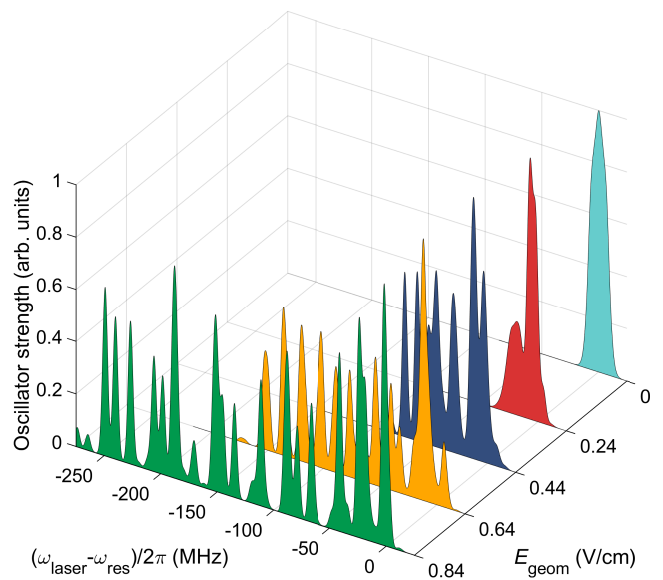}
\caption{ \label{fig:Spectra1} 
The normalized oscillator strength for the Ca$^+(3D_{3/2} \rightarrow 52F_{5/2})$ resonant transition at various $E_{\rm geom}$. Correspondingly, $\alpha=8.52\times 10^{6}$~V/m$^{2}$ , $\beta=3.32\times 10^{4}$~V/m$^{2}$, and $\Omega_{\rm rf}/2\pi = 3.5$~MHz. Up to $\pm 12$ photons in absorption and emission are included in the Floquet calculations for convergence. The Gaussian convolution is performed on the calculated results by considering a~5-MHz laser linewidth.}
\end{figure}

\subsection{Trap-induced Rydberg spectra}
We investigate the spectroscopic features of the Ca$^{+}(3D_{3/2})+h\nu$ $\rightarrow$ Ca$^{+}(nF_{5/2})$ transition line when a single Ca$^{+}$ ion is confined in a Paul trap. We start with $E_{\rm geom}=0$ in Eq.~(\ref{eqn_Hp}). The coupling of the electron to the linear Paul trap, Eq.~(\ref{eqn_ePhi}), is considered with the rf ($\alpha=8.52\times 10^{6}$~V/m$^{2}$) and the electrode ($\beta=3.32\times 10^{4}$~V/m$^{2}$) field gradients. Multiphoton absorption and emission Floquet transitions, i.e., Ca$^{+}(3D_{3/2})+qh\nu$ $\rightarrow$ Ca$^{+}(nF_{5/2})$, with up to $q=12$ photons absorbed and emitted, are considered.

\begin{figure}
\includegraphics[width=8.4cm]{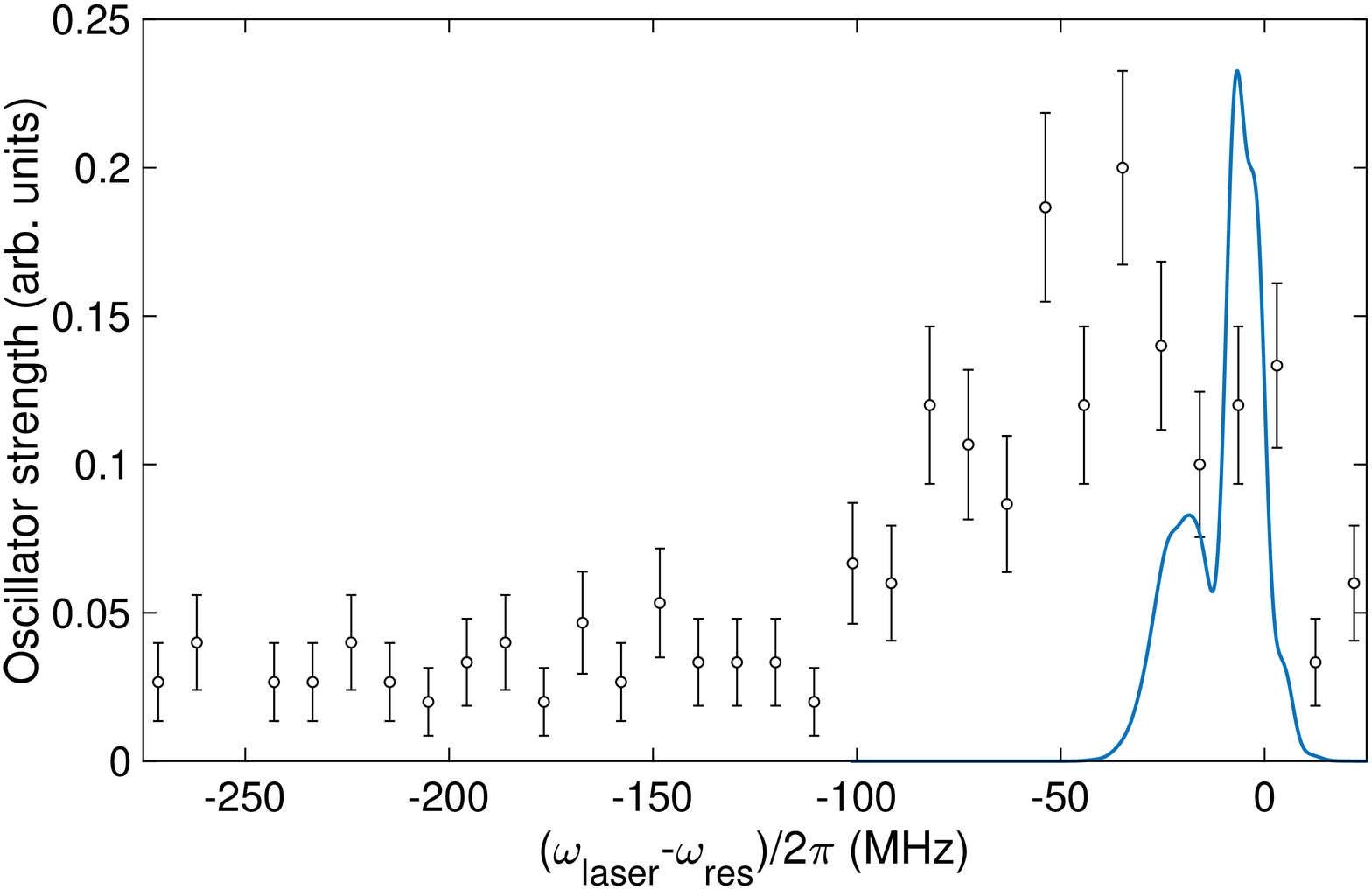}\\
\vspace{0.2cm}
\includegraphics[width=8.4cm]{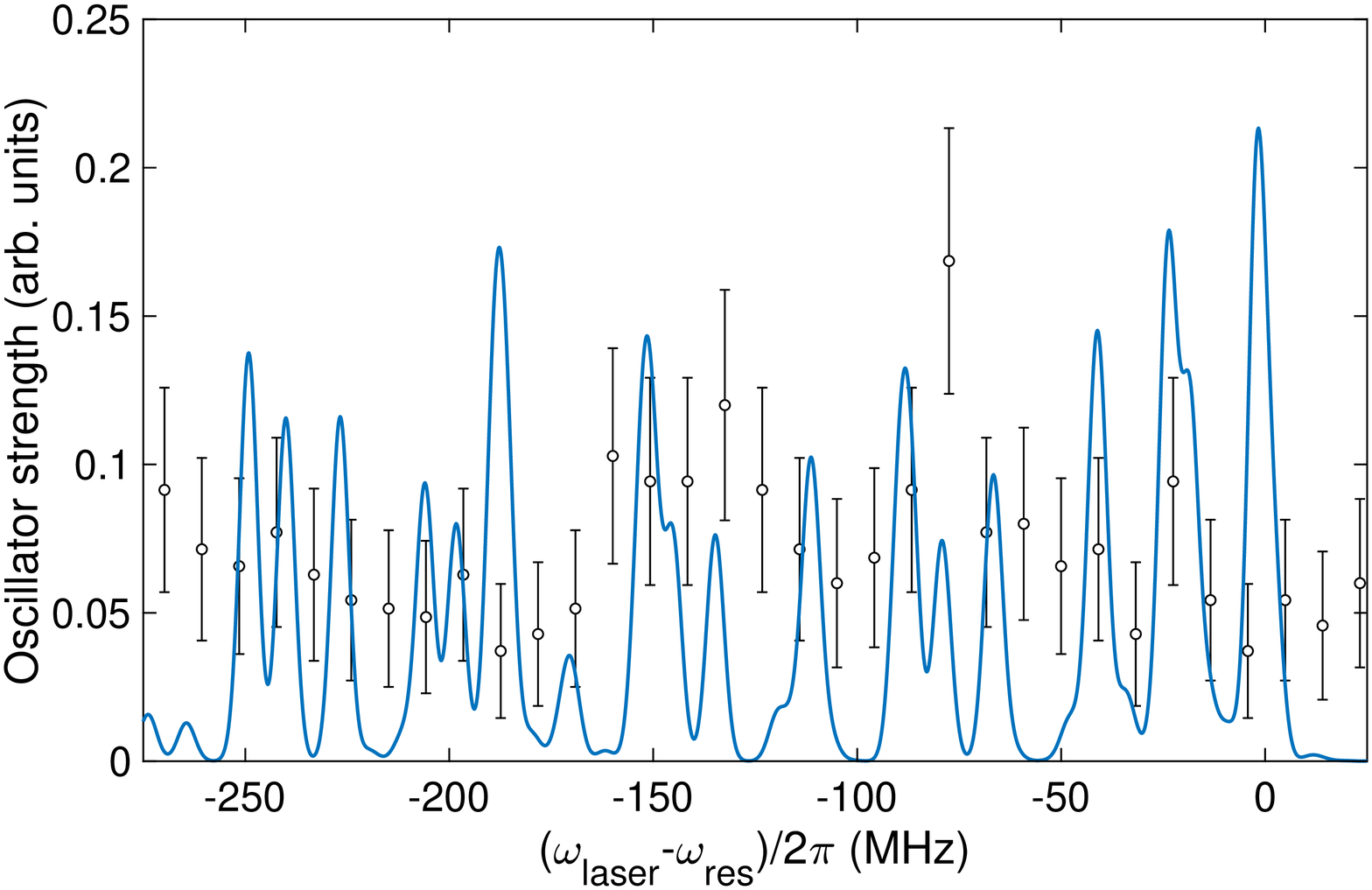}
\caption{ \label{fig:Spectra2} 
The oscillator strength for the Ca$^+(3D_{3/2} \rightarrow 52F_{5/2})$ resonant transition with $E_{\rm geom}=0.24$ V/cm (upper panel), and $E_{\rm geom}=0.84$ V/cm (lower panel).  The transverse and longitudinal trap frequencies are $\omega_{\rm radial}/2\pi=200$ kHz and $\omega_{\rm axial}/2\pi=90$ kHz, respectively. The parameters used are the same as those in Fig.~\ref{fig:Spectra1}. Experimental data are courtesy of the Mainz group~\cite{Mainz2019} and the error bars depict the quantum projection noise. The Gaussian convolution is performed on the calculated results by considering a~5-MHz laser linewidth.}
\end{figure}

\begin{figure}
\includegraphics[width=8.4cm]{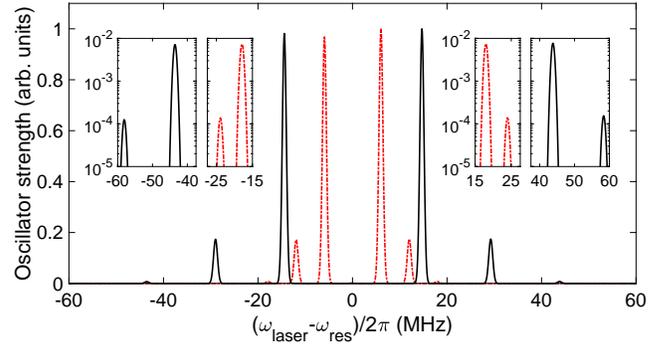}
\caption{ \label{fig:Spectra23P} 
The normalized oscillator strength for the Ca$^+(3D_{3/2} \rightarrow 23P_{1/2})$ resonance transition with $E_{\rm geom}=1.6$~V/cm, $\Omega_{\rm rf}/2\pi=14.56$~MHz, $\alpha=3.161\times 10^{8}$~V/m$^2$, and $\beta=1.286\times 10^{6}$~V/m$^2$ (solid black line) and $E_{\rm geom}=0.1$~V/cm, $\Omega_{\rm rf}/2\pi=5.98$~MHz, $\alpha=1.298\times 10^{8}$~V/m$^2$, and $\beta=1.286\times 10^{6}$~V/m$^2$ (dashed--dotted red line). These results in black and red should be compared with the experimental spectra in Figs.~2(b) and 2(c) in Ref.~\cite{Mokhberi2019}, respectively. In particular, the formation of Rydberg side bands at about $\pm 45$ and $\pm 55$ MHz are visible in the inset plots and detected in the experiment. A~1-MHz laser linewidth is used for the Gaussian convolution of the calculated spectra.}
\end{figure}

Figure~\ref{fig:Mollow_Triplet} shows the Mollow triplets formed near the Ca$^+(52F)$ Rydberg line, for two experimental rf frequencies, $\Omega_{\rm rf}/2\pi=3.5$ and 5.2~MHz~\cite{SchmidtKaler2015} with different Floquet channels (1-photon, 2-photon, and 12-photon absorption and emission). In the lower panel of Fig.~\ref{fig:Mollow_Triplet}, the separation between the two outer peaks is $2\Omega_{\rm rf}$, indicating the convergence of the results. Note that the calculated oscillator strengths are convoluted with a Gaussian laser linewidth of 5~MHz full width at half maximum. As this linewidth is greater than the rf frequency, $\Omega_{\rm rf}/2\pi=3.5$~MHz (solid blue curve), only one broad maximum is visible. The coupling in the Mollow triplet is due to the electron--trap interaction, Eq.~(\ref{eqn_ePhi}). The $\hat H_{\rm trap}^{\rm rf}$ matrix elements obey the $\Delta l=0,2$ and $\Delta m=\pm 2$ selection rules, while the $\hat H_{\rm trap}^{\rm dc}$ matrix elements select the $\Delta l=0,2$ and $\Delta m=0$ transitions.

The effect of the residual electric field, $E_{\rm geom}$, on the spectral line, $3D_{3/2} \rightarrow 52F_{5/2}$, is examined in Fig.~\ref{fig:Spectra1}. The calculated oscillator strengths for this transition are shown for different values of the residual electric field, $0\leq E_{\rm geom} \leq 0.84$~V/cm. The transition matrix elements are integrated over the ground motional state of the trap, e.g., Eq.~(\ref{eq:motional}). Spectral convolution is carried out by a Gaussian function with the 5-MHz laser linewidth. In Fig.~\ref{fig:Spectra2}, the calculated oscillator strengths are compared with the observed spectra~\cite{SchmidtKaler2015,footnote_about_51F}, at $E_{\rm geom}=0.24$ and $0.84$~V/cm. As observed in the experiment, the calculated line shape of the resonance confirms the strong state-dependent coupling to the static and oscillatory electric field potentials in the trap.

Figure~\ref{fig:Spectra23P} presents the calculated oscillator strength for the Ca$^+(3D_{3/2}\rightarrow 23P_{1/2})$ transition for two sets of trap parameters, as in Ref.~\cite{Mokhberi2019}: $E_{\rm geom}=1.6$~V/cm, $\Omega_{\rm rf}/2\pi=14.56$~MHz, $\alpha=3.161\times 10^8$~V/m$^2$, and $\beta=1.286\times 10^6$~V/m$^2$; and $E_{\rm geom}=0.1$~V/cm, $\Omega_{\rm rf}/2\pi=5.98$~MHz, $\alpha=1.298\times 10^8$~V/m$^2$, and $\beta=1.286\times 10^6$~V/m$^2$. The results are convoluted with a Gaussian function of 1-MHz laser linewidth. The peaks, including small far-detuned bumps, agree almost perfectly with the maxima in the experimental spectra in Figs.~2(b) and 2(c) in Ref.~\cite{Mokhberi2019}.

\section{Summary and Outlook}
This work is a description of the first fully variational calculation of the Rydberg spectra of a single ion in a Paul trap. All relevant coupling terms in the Hamiltonian of the ion in the trap are accounted for. The time-periodic rf field is treated nonperturbatively within the Floquet formalism. The motional state of the ion in the trap is also considered. The quantum defect parameters and static and tensor dipole polarizabilities for highly excited states of Ca$^+$ are obtained and compared with available measurements. Precise trapped-induced Rydberg ion [Ca$^+(52F)$ and Ca$^+(23P)$] spectra are calculated. These spectra with their sensitivity to trap or external static and time-varying fields can be used as exquisite probes of residual and stray electric field fluctuations near electrode surfaces and for quantum nonequilibrium dynamics of ion qubits. The extremely large polarizabilities of and controlled long-range interactions between Rydberg states can be employed for ion imaging~\cite{Gross2020}. Future studies of qubit operation, fidelity, and fast computation with trapped Rydberg ions should benefit from such spectral analysis.

\section*{Acknowledgments}
We are grateful to the Mainz group (Schmidt-Kaler and Mokhberi) for extremely valuable discussions and access to the experimental data used here. M.P. thanks the National Science Centre, Poland, for financial support under Grant No. 2017/01/X/ST4/00326. H.R.S. acknowledges the support from the NSF  through a grant for ITAMP at Harvard University.

\setcounter{equation}{0} 
\renewcommand{\theequation}{A.\arabic{equation}}

\section*{Appendix}
Matrix elements of the terms of the time-independent Floquet--Hamiltonian, presented in Eq.~(\ref{Floquet-Matrix}), with the basis set  $\{\xi_\eta\}$, Eq.~(\ref{basis_set_xi}), are explicitly given below:
\begin{widetext}
\begin{eqnarray}
\left [ \mathbf{E} \right ]_{\eta^\prime,\eta} 
&=& E_{n,l,j} \delta_{n',n}\delta_{l',l}\delta_{m',m} \delta_{k_X^{\prime},k_X} \delta_{k_Y^{\prime},k_Y} \delta_{k_Z^{\prime},k_Z}, 
\end{eqnarray}
\begin{eqnarray}
\left [ \mathbf{H}_{\rm trap}^{\rm dc} \right ]_{\eta^\prime,\eta} 
&=& \beta \langle \xi_{\eta^\prime}|x^2+y^2-2z^2|\xi_{\eta}\rangle  \nonumber \\
&=&\beta \langle \psi_{n'}|r^2|\psi_n \rangle \left (\delta_{l',l}\delta_{m',m}-3\langle Y_{l',m'}|\cos^2\theta|Y_{l,m} \rangle \right )  \delta_{k_{X}^{\prime},k_X} \delta_{k_{Y}^{\prime},k_Y}  \delta_{k_{Z}^{\prime},k_Z}, 
\end{eqnarray}
\begin{eqnarray}
\left [ \mathbf{H}_{\rm Ie}^{\rm dc} \right ]_{\eta^\prime,\eta} 
&=& 2\beta \langle \xi_{\eta^\prime} |xX+yY-2zZ| \xi_{\eta} \rangle  \nonumber \\
&=& 2\beta \langle \psi_{n'}|r| \psi_{n} \rangle \left ( \langle Y_{l',m'} |\sin\theta\cos\phi|Y_{l,m} \rangle 
\langle \psi_{k_X^{\prime}}|X| \psi_{k_X}\rangle \delta_{k_{Y}^{\prime},k_{Y}} \delta_{k_{Z}^{\prime},k_{Z}} \right . \nonumber \\
&& +  \langle Y_{l',m'} |\sin\theta\sin\phi|Y_{l,m} \rangle \langle \psi_{k_Y^{\prime}}|Y|\psi_{k_Y} \rangle \delta_{k_{X}^{\prime},k_{X}} \delta_{k_{Z}^{\prime},k_{Z}} \nonumber \\
&& \left . - 2\langle Y_{l',m'} |\cos\theta|Y_{l,m} \rangle \langle \psi_{k_Z^{\prime}}|Z| \psi_{k_Z}\rangle \delta_{k_{X}^{\prime},k_{X}}  \delta_{k_{Y}^{\prime},k_{Y}} \right ), 
\end{eqnarray}
\begin{eqnarray}
\left [ \mathbf{H}_{\rm I} \right ]_{\eta^\prime,\eta}
&=& \delta_{n^\prime,n}\delta_{l^\prime,l}\delta_{m^\prime,m} \sum_{\rho=X,Y,Z} \omega_{\rho} \left(k_\rho+\frac{1}{2} \right) \delta_{k_{\rho}^{\prime},k_\rho}, 
\end{eqnarray}
\begin{eqnarray}
\left [ \mathbf{I} \right ]_{\eta^\prime,\eta} 
&=& \delta_{n',n}\delta_{l',l}\delta_{m',m} \delta_{k_X^{\prime},k_X} \delta_{k_Y^{\prime},k_Y} \delta_{k_Z^{\prime},k_Z}, 
\end{eqnarray}
\begin{eqnarray}
\left [\mathbf{H}_{\rm trap}^{\rm rf} \right ]_{\eta^\prime,\eta}
&=& -\alpha \langle\xi_{\eta^\prime} |x^2-y^2|\xi_{\eta} \rangle  \nonumber \\
&=& -\alpha\langle \psi_{n'} |r^2| \psi_{n}\rangle \left (\langle Y_{l',m'} |\sin^2\theta\cos^2\phi|Y_{l,m}\rangle \right . \nonumber\\
&& \left . -\langle Y_{l',m'}|\sin^2\theta\sin^2\phi |Y_{l,m}\rangle \right ) \delta_{k_X^{\prime},k_X} \delta_{k_Y^{\prime},k_Y} \delta_{k_Z^{\prime},k_Z}, 
\end{eqnarray}
\begin{eqnarray}
\left [\mathbf{H}_{\rm Ie}^{\rm rf} \right ]_{\eta^\prime,\eta}
&=& -2\alpha \langle\xi_{\eta^\prime} |xX-yY| \xi_\eta \rangle  \nonumber \\
&=& -2\alpha \langle \psi_{n'} |r| \psi_{n}\rangle \left (\langle Y_{l',m'} |\sin\theta\cos\phi|Y_{l,m}\rangle \langle  \psi_{k_X^{\prime}} |X| \psi_{k_X} \rangle \delta_{k_Y^{\prime},k_Y} \delta_{k_Z^{\prime},k_Z}  \right .\nonumber\\
&& \left . -\langle Y_{l',m'}|\sin\theta\sin\phi |Y_{l,m}\rangle  \langle  \psi_{k_Y^{\prime}} |Y| \psi_{k_Y} \rangle \delta_{k_X^{\prime},k_X} \delta_{k_Z^{\prime},k_Z}  \right ), 
\end{eqnarray}`
\begin{eqnarray}
\left [ \mathbf{H}_{\rm geom} \right ]_{\eta^\prime,\eta} 
&=& E_{\rm geom}\langle \xi_{\eta^\prime}|z |\xi_{\eta}\rangle \nonumber \\
&=& E_{\rm geom}\langle \psi_{n'} |r| \psi_n \rangle  \langle Y_{l',m'}|\cos\theta | Y_{l,m}\rangle \delta_{k_{X}^{\prime},k_{X}} \delta_{k_{Y}^{\prime},k_{Y}} \delta_{k_{Z}^{\prime},k_{Z}},
\end{eqnarray}
\end{widetext}
where $\eta$ denotes a~superindex containing quantum numbers $\{n,l,m,k_X,k_Y,k_Z\}$. Simple expressions for the above angular matrix elements are reported in the Supporting Information in Ref.~\cite{Pawlak2017}.

\end{document}